\documentclass[aps,prl,twocolumn,superscriptaddress,showpacs,amsmath]{revtex4}
\usepackage{graphicx}

\begin{document}
\title{Theory of domain structure in ferromagnetic phase of diluted
magnetic semiconductors near the phase transition temperature}

\author{Vladimir Stephanovich}
\homepage{http://cs.uni.opole.pl/~stef}
\email{stef@math.uni.opole.pl} \affiliation {Opole University,
Institute of Mathematics and Informatics, Opole, 45-052, Poland}
\date{\today }

\begin{abstract}
We discuss the influence of disorder on domain structure formation
in ferromagnetic phase of diluted magnetic semiconductors (DMS) of
p-type. Using analytical arguments we show the existence of
critical ratio $\nu _{\rm {cr}}$ of concentration of charge
carriers and magnetic ions such that sample critical thickness
$L_{\rm{cr}}$ (such that at $L<L_{\rm{cr}}$ a sample is
monodomain) diverges as $\nu \to \nu _{\rm {cr}}$. At $\nu > \nu
_{\rm {cr}}$ the sample is monodomain. This feature makes DMS
different from conventional ordered magnets as it gives a
possibility to control the sample critical thickness and emerging
domain structure period by variation of $\nu $. As concentration
of magnetic impurities grows, $\nu _{\rm {cr}}\to \infty$
restoring conventional behavior of ordered magnets. Above facts
have been revealed by examination of the temperature of transition
to inhomogeneous magnetic state (stripe domain structure) in a
slab of finite thickness $L$ of p-type DMS. Our analysis is
carried out on the base of homogeneous exchange part of magnetic
free energy of DMS calculated by us earlier [\prb, {\bf 67},
195203 (2003)].
\end{abstract}

\maketitle

The structure of domains and domain walls in conventional ordered
magnets has been well studied both experimentally and
theoretically several decades ago (see, \cite{land8,TCD} and
references therein). On the other hand, the important question
about the domain structure formation in the ferromagnetic phase of
p-doped diluted magnetic semiconductors (DMS)\cite{DMS,DMS1}
(which can be regarded as disordered magnets) has not been
addressed yet. The characteristics of the domain structure in the
DMS films of III-V type have been investigated in Ref
\cite{domains} (see also references therein). In our opinion, the
main problem was the lack of suitable "continuous" free energy
functional of DMS, which is necessary to theoretically study their
domain structure. Recently \cite{SS2} such free energy functions
have been derived microscopically from Ising and Heisenberg models
for DMS. In the language of phenomenological theory of magnetism
these functions correspond to so-called homogeneous exchange parts
of total phenomenological free energy of DMS. To describe the
domain structure properties, these contributions should be
completed by inhomogeneous exchange and magnetic anisotropy
energies. It was demonstrated experimentally (see \cite{DMS1} and
references therein) that magnetic anisotropy exists in DMS of
(Ga,Mn)As type. At the same time it was demonstrated in
\cite{DMS1} that unstrained samples (which can be well described
by Heisenberg model \cite{SS2}) have easy plane magnetic
anisotropy, while uniaxially strained samples (Ising model
\cite{SS2}) have anisotropy of easy axis type.

It is well known (see, e.g. \cite{land8}) that at low temperatures
the domain pattern formation is primarily due to the rotation of
magnetization vector with constant modulus being saturation
magnetization $M_0$. On the contrary, for the temperatures close
to $T_c$, this structure is formed by the variation of modulus of
${\vec M}$ rather then its rotation. This means that above
homogeneous exchange part of the magnetic energy of DMS will only
contribute to its domain structure in the vicinity of $T_c$. At
low temperatures the influence of disorder on domain structure of
DMS will be small so that it will resemble very much the domain
structure of conventional magnetically ordered substances.

In the present paper we suggest a theory of inhomogeneous magnetic
state (stripe domain structure) in the DMS slab in the vicinity of
ferromagnetic phase transition temperature. We analyze the sample
of finite thickness $L$. We show that the impurity character of
ferromagnetism in DMS results in substantial narrowing of the
region of temperatures and sample thicknesses, when domain
structure exists. For example, beyond mean field approximation
(i.e. when disorder in magnetic ions subsystem becomes
substantial) even at zero temperature domain structure appears not
at $L=0$ (as in the case of ordered magnet, see \cite{TCD,bariv}),
but at some threshold value $L_{\rm{tr}}$, depending on the ratio
$\nu $ of charge carriers and magnetic ions concentrations, see
Fig.1. This effect makes the domain structure of DMS (disordered
magnets) qualitatively different from that of conventional ordered
magnets. The developed formalism can be easily applied for thin
DMS films.

Consider the slab-shaped sample of DMS with slab thickness L
(Fig.1,lower panel). Let $z$ axis is magnetic anisotropy axis (and
$xy$ plane is the plane of anisotropy for Heisenberg model). The
phenomenological free energy of DMS near $T_c$ can be written in
the form (see, e.g. \cite{TCD})
\begin{equation}
F = \int dv\left\{ \frac{1}{2}\alpha \left( {\nabla \vec
M}\right)^2  +f_{\rm {AN}}^P+f^P(M)-\frac 12\vec M\vec H_D
\right\}, \label{apr1}
\end{equation}
where ${\vec M}$ is a magnetization vector, $\alpha $ is
inhomogeneous exchange constant, ${\vec H}_D$ is a demagnetizing
field, $P$ stands for $H$ (Heisenberg model) or $I$ (Ising model),
$f_{\rm {AN}}^P$ are the anisotropy energies and $f^P(M)$ are the
homogeneous exchange energies. For Heisenberg model (easy plane
anisotropy)\cite{SS2}
\begin{eqnarray}
&&f_{\rm {AN}}^H=\frac{1}{2}\beta M_x^2,\ \beta >0\label{anizH}\\
&&f^H(m)=\frac 12m^2\left( 1-2A_1^H\right) +\frac 1{20}m^4A_3^H +
...,\label{lanH}\\
&&A_n^H =\int_0^\infty {\cal B}_{1/2}^H(\pi t)e^{-{\cal F}_0(t/2T)}{\cal F}%
_1^n\left( \frac t{2T}\right) dt,\label{AnH}
\end{eqnarray}
where $m=|{\vec m}|$, ${\vec m}={\overline {<\vec S>}}/S$, $S$ is
a spin of a magnetic ion. Bar means the averaging over spatial
disorder in magnetic ion subsystem in DMS, while angular brackets
mean the thermal averaging, see \cite{SS} for details. The
relation between $m$ and $M$ in this case is usual: $m$=$M/M_0$,
where $M_0={\overline {<\vec S>}}(T=0)$ is saturation
magnetization.

For Ising model (easy axis anisotropy)
\begin{eqnarray}
&&f_{\rm {AN}}^I=\frac{1}{2}\beta
\left(M_x^2+M_y^2\right),\ \beta >0,\label{anizI}\\
&&f^I(m)=\frac 12m^2\left( 1-A_1^I\right) +\frac
1{24}m^4A_3^I+...,\label{lanI}\\
&&A_n^I =\int_0^\infty {\cal B}_{1/2}^I(\pi t)e^{-{\cal F}_0(t/2T)}{\cal F}%
_1^n\left( \frac t{2T}\right) dt.\label{AnI}
\end{eqnarray}
Here
\begin{eqnarray}
&&{\cal B}_{1/2}^H(x)=\frac{1+x\coth x}{3\sinh x} ,
\ {\cal B}_{1/2}^I(x)=\frac 1{\sinh x},\label{bHI}\\
&&{\cal F}_{0}(x)+i{\cal F}_{1}(x)= \int_{V}n({\vec{r}})
\left[1-e^{-i J(\vec{r})x}\right]d^3r,\label{f0}
\end{eqnarray}
${\cal F}_{0,1}$ are the real and imaginary parts of Fourier image
of distribution function of random magnetic fields, acting among
magnetic impurities in DMS, $J(\vec{r})$ is an interaction between
spins of magnetic impurities in the Heisenberg or Ising
Hamiltonians (this is actually RKKY interaction, see \cite{SS,SS2}
for details), $n({\vec{r}})$ is (spatially inhomogeneous)
concentration of magnetic ions. Index 1/2 in the functions
(\ref{bHI}) means that free energies (\ref{lanH}) and (\ref{lanI})
have been derived for spin 1/2 (see \cite{SS2}). However, our
analysis shows that all our results remain qualitatively the same
for arbitrary spin.

Equilibrium distribution of magnetization in DMS can be obtained
from the equation of state ${\delta F}/{\delta {\vec m}}=0$ and
Maxwell equations ${\rm {rot}}\ {\vec h}=0$, ${\rm div}\
\left({\vec h}+4\pi {\vec m}\right)=0$ with boundary conditions
for slab geometry
\begin{eqnarray}
&&\left.h_{x}\right|_{z=\pm\frac
L2}=\left.h_{x}^{(e)}\right|_{z=\pm\frac L2},\
 \left.\frac{\partial {\vec m}}{\partial z}\right|_{z=\pm\frac
L2}=0, \nonumber \\
&&\left[h_{z}+4\pi m_z\right]_{z=\pm\frac
L2}=\left.h_{z}^{(e)}\right|_{z=\pm\frac L2},\nonumber \\
\label{GranUsl}
\end{eqnarray}
where ${\vec h}={\vec H}_D/M_0$, ${\vec h}^{(e)}$ is a
demagnetizing field in a vacuum. It was shown in \cite{TCD} that
for sufficiently thick slabs \cite{thick} and $\beta <4\pi $ the
following equation for distribution of magnetization in DMS can be
derived from the above equation of state with respect to
(\ref{apr1})
\begin{equation}\label{master}
\mu _{\perp }\frac{\partial ^{2}}{\partial x^{2}}\left( \alpha
\frac{\partial ^2m_z}{\partial x^2}-b_Pm_{z}-c_P m_{z}^{3}\right)
-4\pi \frac{\partial ^{2}m_{z}}{\partial z^2}=0,
\end{equation}
where $\mu _\perp=1+4\pi /\beta$, $b_H=1-2A_1^H$, $c_H=A_3^H/5$,
$b_I=1-A_1^I$, $c_I=A_3^I/6$, where in the ferromagnetic phase
($T<T_c$) functions $b_{\rm P }(T)<0$ ($P=H,I$). It should be
noted here, that the different forms of anisotropy energies for
Heisenberg (\ref{anizH}) and Ising (\ref{anizI}) models in the
above suppositions do not influence the form of equation
(\ref{master}).

It was shown in \cite{TCD} (see also \cite{bariv}), that
transition from paramagnetic phase to the ferromagnetic phase with
domain structure (domain state) occurs via phase transition of
second kind. This means that for the determination of transition
temperature $T_k$ to the domain state it is sufficient to consider
the linearized version of Eq. (\ref{master}). We look for its
solution in the form
\begin{equation}\label{sol}
m_z=A\cos qz\cos kx.
\end{equation}
Here $\cos qz$ determines the spatial inhomogeneity along $z$
direction, while $\cos kx$ defines a "linear" domain structure (in
$x$ direction so that domain walls lie in $yz$ plane) with a
period $d=2\pi /k$.

Substitution of solution (\ref{sol}) into the linearized version
of (\ref{master}) gives the equation relating $q$ and $k$
\begin{equation}\label{qk1}
q^2=\frac{\mu _\perp}{4\pi}k^2\left(\zeta _P-\alpha k^2 \right),\
\zeta _P=-b_P
\end{equation}
To obtain the dependence of $T_k$ on sample thickness, we need one
more equation relating $\zeta $ (and by its virtue $T_k$, see
(\ref{AnH}),(\ref{AnI})), $k$ and $q$. Such equation can be gotten
substituting (\ref{sol}) into boundary conditions (\ref{GranUsl})
with respect to vacuum solutions
$h_z^{(e)}=C\exp\left(-k|z|\right)\cos kx$,
$h_x^{(e)}=C\exp\left(-k|z|\right)\sin kx$. It reads
\begin{equation}\label{second}
\tan \frac{qL}{2}=\frac{\mu _\perp k}{q}.
\end{equation}
Equations (\ref{qk1}) and (\ref{second}) constitute a close set of
equations for the instability temperature $T_k$ and equilibrium
domain structure period as a function of sample thickness $L$ and
concentration ratio $\nu $. The equations (\ref{qk1}) and
(\ref{second}) can be reduced to a single equation for the
dependence of $b_P$ on $y=k\sqrt{\alpha }$
\begin{equation}\label{single}
y\sqrt {\frac{\mu _\perp}{4\pi }\left( \zeta _P - y^2 \right)}=\pi
n+\frac{2}{\eta}\arctan\sqrt{\frac{4\pi \mu _\perp}{\zeta _P -
y^2}},
\end{equation}
where $\eta ^2=L^2/\alpha $. The equation (\ref{single}) is a
single equation for $\zeta (y)$ at different $\eta $. This
dependence has the form of a curve with a minimum. The real
transition to domain state in DMS occurs when $T_k$ reaches its
maximal value as a function of $y$. This can be demonstrated by
substitution of a solution of nonlinear equation (\ref{master}) in
the form of infinite series in small parameter proportional to
$|T-T_k|$ into free energy (\ref{apr1}) with its subsequent
minimization over $y$ similarly to Ref.\cite{TCD}. Since $\zeta
_P=-b_P$ for both models is a decreasing function of temperature
(e.g. for both models in a mean field approximation $\zeta =1/\tau
-1$, $\tau =T/T_{cMF}$ \cite{SS,SS2}), the coordinates of minimum
$\zeta _P^{\rm {min}}$ and $y^{\rm {min}}$ of the curve $\zeta
(y)$ determine the equilibrium temperature of a phase transition
to the domain state $T_k$ and equilibrium period of emerging
domain structure $\lambda =2\pi /y^{\rm {min}}$ as functions of
dimensionless sample thickness $\eta $.
\begin{figure}[th]
\vspace*{-5mm} \hspace*{-5mm} \centering{\
\includegraphics[width=8cm]{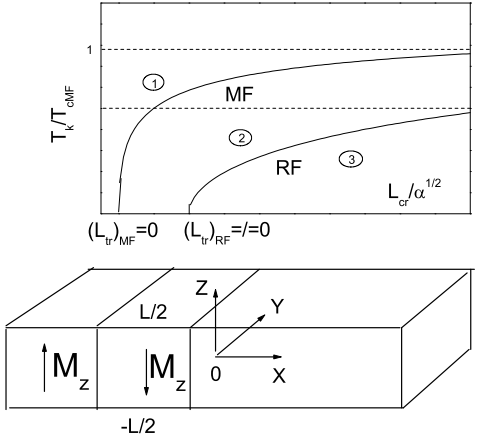}}
\caption{Upper panel - schematic phase diagram of the DMS slab in
the coordinates temperature - thickness. Region 1 corresponds to
paramagnetic phase for both mean field (MF) and random field (RF)
approximations, region 2 - paramagnetic phase for MF and domain
state for RF, region 3 - domain state for both cases. The nonzero
threshold sample thickness occurs for RF case. Horizontal
asymptotes correspond to $\tau _c(\nu )=T_c/T_{cMF}$, $T_c$ is a
temperature of phase transition into ferromagnetic homogeneous
(i.e. domainless) state. Lower panel - geometry of the sample.}
\end{figure}

We have taken the minimum of implicit function $\zeta (y)$
(\ref{single}) numerically to get the dependencies $\lambda (\eta
)$ and $\zeta _P^{\rm {min}}(\eta ) $. They are reported on inset
to Fig.2. It is seen that dependence $\zeta _P^{\rm {min}}(\eta )
$ decays rapidly as $\eta \to \infty$, at $\eta \to 0$ $ \zeta
_P^{\rm {min}} \to \infty$. It can be shown that at large $\eta $
$\lambda \propto \sqrt{\eta }=\sqrt{L/\sqrt{\alpha }}$. This
behavior is typical for stripe domain structure in ordered magnets
(see \cite{TCD,bariv}) and is seen on the figure.

Now we consider the explicit dependencies $b_P(\nu ,\tau )$,
$\nu=n_c/n_i=x_c/x_i$, $x_{c,i}=n_{c,i}\Omega $, $\Omega $ is DMS
unit cell volume, $n_c$ and $n_i$ are charge carriers (electrons
or holes) and magnetic ions concentrations respectively. We have
in dimensionless variables
\begin{eqnarray}
&&\zeta _P=\frac{\tau }{36\pi ^2\nu ^2}\int _0^\infty {\cal
B}_{1/2}^P\left(\frac{\xi \tau}{12\nu
}\right)\Phi (\xi )d\xi -1,\label{a1h}\\
&& \Phi (\xi )=\exp\left[\frac{\varphi _0(\xi ) }{6\pi \nu
}\right]\varphi _1(\xi ),\quad P=H,I,\nonumber
\end{eqnarray}
where $\xi =J_0x_c^{4/3}t/2T$. In the expressions (\ref{a1h}) we
used RKKY potential in the simplest possible form corresponding to
one-band carrier structure
\begin{equation}\label{RKKY}
J({\vec r})=-J_0x_c^{4/3}R(2k_fr),\ F(x)=\frac{x\cos x-\sin
x}{x^4},
\end{equation}
where $J_0=\left( 3/\pi \right) ^{1/3} \left(3/2\hbar
^2\right)J_{ci}^{2}\Omega ^{2/3}m_d$, $J_{ci}$ is a carrier-ion
exchange constant, $m_d$ is the density of states effective mass.
Functions $\varphi _0(\xi )$ and $\varphi _1(\xi)$ have the form
\[
\varphi _{0}\left( \xi \right)+i\varphi _{1}\left( \xi \right)
=\int_{0}^{\infty }\left\{ 1-\exp \left( -i\xi F(y)\right)
\right\} y^{2}dy.
\]
The dependencies $\zeta _H(\tau )$ and $\zeta
_I(\tau )$ at different $\nu $ are shown on Fig.2. In a mean field
(MF) approximation $\zeta _{HMF}=\zeta _{IMF}=\zeta _{MF}=1/\tau
-1$ is unbounded at $T=0$, while beyond this approximation
functions $\zeta _H$ and $\zeta _I$ have finite values at $T=0$
\begin{equation}
\zeta _{H0}=\frac 23 \zeta _{I0}=\frac{2}{9\pi ^2\nu
}\int_0^{\infty}\Phi (\xi )\frac{d\xi }{\xi }-1.\label{crit}
\end{equation}
\begin{figure}[th]
\vspace*{-6mm} \hspace*{-5mm} \centering{\
\includegraphics[width=8.5cm]{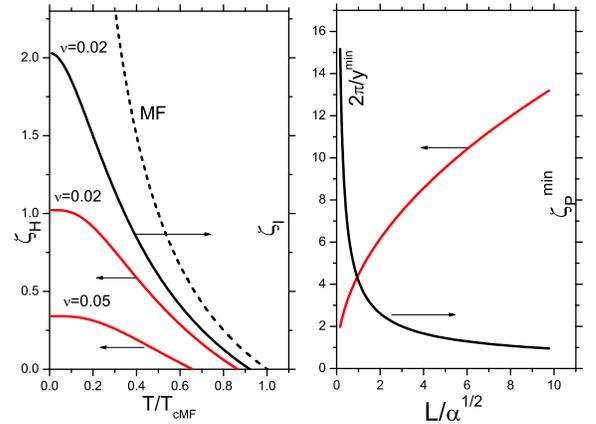}}
\vspace*{-6mm} \caption{Left panel: dependence of $\zeta _H$ and
$\zeta _I$ on normalized temperature at different $\nu $. Points
where $\zeta _P=0$ correspond to $\tau _c(\nu)$. Dashed line
labelled MF corresponds to $\zeta _{MF}=1/\tau -1$. Right panel:
equilibrium period of emerging domain structure and parameter
$\zeta $ versus dimensionless sample thickness at $\mu _\perp=10$
.}
\end{figure}

These finite values are indeed "responsible" for the emergence of
threshold sample thickness $L_{\rm {tr}}$ beyond mean field
approximation. Having dependencies (\ref{a1h}), we can solve them
numerically for $\zeta _P^{\rm {min}}$ (determined above from
(\ref{single})) to obtain the dimensionless phase transition
temperature $\tau _k=T_k/T_{cMF}$ as a function of critical sample
thickness $L_{\rm {cr}}/\sqrt{\alpha }$. In other words, here we
have the phase diagram of DMS slab in the coordinates $(T,L)$.
This is reported on Fig.3.

\begin{figure}[th]
\vspace*{-6mm} \hspace*{-5mm} \centering{\
\includegraphics[width=8.5cm]{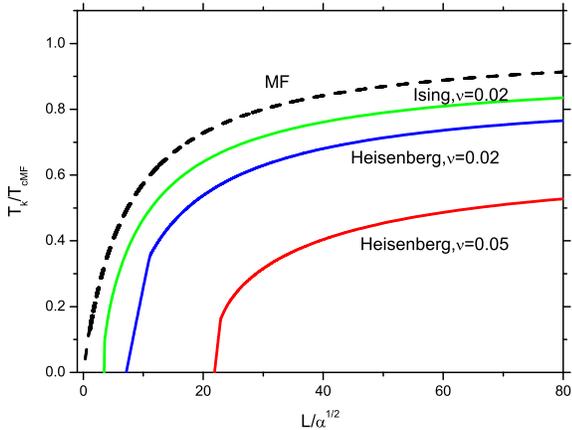}}
\vspace*{-6mm} \caption{Phase diagram of DMS slab in the
coordinates $T,L$. Horizontal asymptotes - $\tau _c(\nu )$ similar
to Fig.1, $\mu _\perp=10$. }
\end{figure}

The presence of $L_{\rm {cr}}(\nu )$ is clearly seen. The
asymptotes for large $L_{\rm {cr}}$ are due to the dependence of
equilibrium (i.e. to the ferromagnetic phase without domain
structure) phase transition temperature $\tau _c=T_c/T_{cMF}$ on
the concentration ratio $\nu $. Latter dependence is given by the
conditions $\zeta _H=0$ and $\zeta _I=0$ (see (\ref{lanH}),
(\ref{lanI})) for Heisenberg and Ising models respectively. It was
shown in Ref. \cite{SS}that impurity ferromagnetism in DMS is
possible for $0<\nu <\nu _{\rm {cr}}^P$ ($P=H,I$, $\nu _{\rm
{cr}}^H=0.0989$, $\nu _{\rm {cr}}^I=0.2473$ \cite{SS}) so that
$\tau _c(\nu _{\rm{cr}})=0$. This means that $\tau _c$ decays as
$\nu $ grows and the region $(T,L)$, where ferromagnetic domain
state exists in DMS, diminishes substantially (compared to the
case of ordered ferromagnets) and vanishes as $\nu \to \nu _{\rm
{cr}}$.

\begin{figure}[th]
\vspace*{-6mm} \hspace*{-5mm} \centering{\
\includegraphics[width=8.5cm]{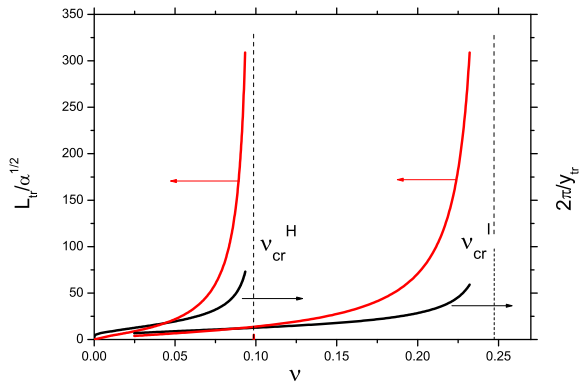}}
\vspace*{-6mm} \caption{Threshold sample thickness (i.e. that at
$T=0$) and corresponding domain structure period as functions of
concentration ratio $\nu $ at $\mu _\perp=10$. Critical ratios for
Heisenberg $\nu _{\rm {cr}}^H$ and Ising $\nu _{\rm {cr}}^I$ are
shown. For mean field approximation (ordered magnet) $\nu _{\rm
{cr}}\to \infty$. }
\end{figure}

Note, that in MF approximation it is very easy to solve
(\ref{a1h}) analytically to get $\tau _{cMF}=1/(\zeta _P+1)$.
Resolving the equations (\ref{crit}) for $\zeta _P^{\rm {min}}$ we
obtain the dependence of threshold thickness on concentration
ratio $\nu $. This dependence along with corresponding period of
domain structure is shown on Fig.4 for Heisenberg and Ising
models. It is seen that as $\nu $ approaches $\nu _{\rm{cr}}$
$L_{\rm{tr}}\to \infty$ so that DMS sample looses its
ferromagnetism (both homogeneous and inhomogeneous). The period of
domain structure also diverges as $\nu \to $ $\nu _{\rm{cr}}$.

This fact makes it possible to control the critical thickness of
DMS sample by changing $\nu $. This, in turn, might give a
possibility to engineer the domain structure in nanocrystals od
DMS, which is useful for many technical applications (see
\cite{domains,DMS1} and references therein). Note that our
formalism permits to calculate $\nu _{\rm{cr}}$ and other
characteristics of domain structure for any temperature (not only
neat $T_c$ and sample geometry).

Now we present some numerical estimations. The major problem here
is uncertain value of inhomogeneous exchange constant $\alpha $.
It can be estimated by the expression \cite{land8} $\alpha \approx
k_BT_ca^2/(M_s\mu _0)$, where $a=4\AA$ is a typical value of
lattice constant for DMS, $M_s \approx 50$ mT \cite{DMS1} is a
saturation magnetization (of localized spin moments) of DMS,
$T_c=T_{cMF}\approx 100 K$ \cite{DMS1} is a temperature of
transition to homogeneous ferromagnetic state in a mean field
approximation, $k_B$ and $\mu _0$ are Boltzmann constant and Bohr
magneton respectively.  Evaluation gives $\alpha ^{1/2}\sim 200
\AA$. From Fig. 4 for $\nu = 0.075$ we have threshold sample
thickness $L_{\rm{tr}}\sim 50 \alpha ^{1/2}=10000 \AA =1\mu $m and
corresponding period of domain structure is $\sim 25 \alpha
^{1/2}=5000 \AA =0.5\mu $m for Heisenberg model. The same values
for Ising model occur at $\nu \sim 0.2$. These values are in fair
agreement with results of Ref. \cite{domains}. Moreover, for
different $\nu $ we have quite different values of $L_{\rm{tr}}$
and $y_{\rm {tr}}$. This is the base for above discussed domain
structure engineering. For our picture to give the quantitative
description of experiment in real DMS, the precise experimental
determination of inhomogeneous exchange constant $\alpha $ and
anisotropy constant $\beta $ is highly desirable.

Here we presented a formalism for the calculation of properties of
domain structure in DMS. Our present results about phase diagram
of DMS is the simplest application of the formalism. Generally, it
permits to calculate all desired properties of domain structure
(like the temperature and concentration dependencies of domain
structure period and domain walls thickness) in the entire
temperature range as well as to account for more complex then slab
sample geometries. Latter can be accomplished by applying
different from (\ref{GranUsl}) boundary conditions. The external
magnetic field can also be easily taken into account. However, far
from $T_k$ the solution of resulting nonlinear differential
equations would require numerical methods.


\begin{thebibliography}{99}
\bibitem{thick} The equations (\ref{lanH}) and (\ref{lanI}) have
been derived for 3D RKKY interaction between localized magnetic
moments in DMS \cite{SS2}. For such consideration to be valid, the
slab thickness $L$ should be more then the range of RKKY
interaction $1/k_f$ ($k_f$ is Fermi wave vector). For DMS, the
typical value of Fermi energy $E_f \sim 0.08$ eV. Hence $1/k_f
=\hbar/\sqrt{2m^*E_f}$, where $m^*\sim 0.1 m_0$ is an effective
mass of charge carrier, $m_0$ is a free electron mass. So, in our
consideration $1/k_f \sim 20 \AA$ which is the lower limit for
plate thickness. However, real plate thickness should be much more
then this threshold value. Namely in the phenomenological theory
of magnetic domains (see, e.g. \cite{TCD} and \cite{bariv}) the
criterion of slab thickness reads $L \gg \alpha ^{1/2}$. For
$\alpha ^{1/2}=200\AA$ both above criteria coinside.
\end{thebibliography}
\end{document}